\documentclass[twocolumn,aps,pre,superscriptaddress,nofootinbib]{revtex4-2}

\usepackage{pifont}

\usepackage{color, colortbl}
\definecolor{Gray}{gray}{0.2}

\usepackage{graphicx}
\usepackage[dvipsnames,svgnames]{xcolor}
\usepackage{bm,bbm}
\usepackage{braket}
\usepackage{amsthm,amsmath,amssymb}
\usepackage{enumerate}
\usepackage{booktabs,multirow}
\usepackage{mathtools,bbding}
\usepackage[percent]{overpic}
\usepackage{microtype}
\usepackage{array,adjustbox,afterpage}
\usepackage[T1]{fontenc}
\usepackage[utf8]{inputenc}
\usepackage{tikz}
\usetikzlibrary{calc}
\usepackage[english]{babel}
\usepackage{newtxtext,newtxmath}

\usepackage{lipsum}

\usepackage[linesnumbered,ruled]{algorithm2e}  %delete ruled and write boxed for box

\makeatletter
\renewcommand{\fnum@figure}{FIG. \thefigure} % Figure --> FIG.
\makeatother

\usepackage[bookmarks=false, colorlinks=true, linkcolor=blue, citecolor=purple, urlcolor=purple]{hyperref}
\usepackage{cleveref}
\usepackage[normalem]{ulem}

\usepackage{lipsum}
\usepackage{makecell}

\Crefname{subfigures}{figure}{figures}
\Crefname{subfigures}{Figure}{Figures}

\newcommand{\xdownarrow}[1]{{\left\downarrow\vbox to #1{}\right.\kern-\nulldelimiterspace}} % for long downarrow ..

\hypersetup{
    breaklinks=true,   % splits links across lines
    colorlinks=true,   % displays links as colored text instead of blocks
    pdfusetitle=true,  % \title and \author values into pdf metadata etc.
}

\begin{document}
\title{Electron delocalization in aromaticity as a superposition phenomenon}
\author{Mahir H. Ye\c{s}iller}
\email{mahirhyesiller@gmail.com}
\affiliation{Faculty of Engineering and Natural Sciences, Kadir Has University, 34083, Fatih, Istanbul, T\"{u}rkiye}
\affiliation{Department of Physics, Ko{\c{c}} University, 34450 Sar{\i}yer, Istanbul, Turkey}
\author{Onur Pusuluk}
\email{onur.pusuluk@gmail.com}
\affiliation{Faculty of Engineering and Natural Sciences, Kadir Has University, 34083, Fatih, Istanbul, T\"{u}rkiye}
\affiliation{Department of Physics, Ko{\c{c}} University, 34450 Sar{\i}yer, Istanbul, Turkey}

%\date{\today}

%***======***%
%  ABSTRACT  %
%***======***%

\begin{abstract}

This letter investigates the applications and extensions of the resource theory of quantum superposition within the realm of quantum chemistry. Specifically, we explore aromaticity, a fundamental concept originally developed to elucidate the structural symmetry, energetic stability, and chemical reactivity of benzene and its derivatives. While aromaticity and its counterpart, antiaromaticity, are associated with the delocalization of electrons between nonorthogonal atomic orbitals, they lack a universally accepted and comprehensive definition. We demonstrate that the genuine quantum superposition exhibited by biorthogonal atomic orbitals effectively captures the aromaticity order of representative monocyclic molecules. These findings reveal that the quantum resource theories hold significant implications, offering fresh insights into our comprehension of chemical bonding phenomena.

\end{abstract}

\maketitle

%***==========***%
%  INTRODUCTION  %
%***==========***%
\textit{Introduction.}--- One counterintuitive concept that makes quantum mechanics an intriguing theory is the superposition principle, which allows quantum systems to exist in multiple states simultaneously~\cite{dirac1981principles}. This exceptional phenomenon has significant implications as a resource for tasks that cannot be accomplished through classical means, exemplified by quantum teleportation~\cite{bennett1993teleporting,hu2023progress}. Therefore, the quantification and manipulation of various forms of quantum superposition, such as quantum coherence and quantum entanglement, have become subjects of great interest~\cite{aberg2006quantifying,baumgratz2014quantifying,streltsov2017colloquium,modi2012classical,adesso2016measures}.

Accordingly, the quantum superposition of distinguishable (orthogonal) states, known as quantum coherence, has a well-grounded resource theory~\cite{baumgratz2014quantifying,streltsov2017colloquium}. When this special form of superposition is shared between systems separated in space, it is referred to as quantum correlation, which is better understood through its quantification and manipulation~\cite{modi2012classical,adesso2016measures}. The basis-independent nature of quantum correlations makes them a crucial resource for many emerging quantum technologies~\cite{dowling2003quantum,georgescu2012quantum,jaeger2018second}. For instance, quantum entanglement, an exemplary manifestation of quantum superposition, is considered a perfect illustration of quantum correlations, and its substantial role as a resource within quantum technologies is widely recognized~\cite{horodecki2009quantum,pezze2018quantum,pirandola2020advances}. Nevertheless, quantum entanglement is only a subset of the most general quantum correlations, known as quantum discord~\cite{ollivier2001quantum,henderson2001classical,modi2010unified,bera2017quantum}.

For indistinguishable (nonorthogonal) states, quantification and manipulation of superposition require generalizing the framework used in coherence theory. This generalization, called the resource theory of superposition (RTS), relaxes the orthogonality condition of basis states to linear independence~\cite{theurer2017resource, torun2021resource, csenyacsa2022golden}. Here, quantum superposition can manifest in two distinct ways: as the linear combination of basis states and through their overlaps~\cite{pusuluk2022unified}. From this perspective, quantum coherence is viewed as a subset of superposition when nonclassicality is formed entirely in the form of a linear combination of basis states, and all the overlaps between these states vanish.

The quantification and manipulation of nonclassicality in optical coherent states can be considered an application of RTS~\cite{theurer2017resource}. However, the importance of overlaps between nonorthogonal states is most evident in molecular electronic states~\cite{pusuluk2022unified}. In the molecular world, nonorthogonal states correspond to mathematical objects representing chemical structures such as atomic orbitals~\cite{pauling2013bonding,helgaker2013molecular}. Chemical bonding and other related quantum chemical phenomena arise from the overlaps between these states~\cite{weinhold1999chemical,weinhold2014hydrogen,frenking2014chemical}. Some of these phenomena form the basis of  (quantum) chemistry science and education~\cite{mcquarrie2008quantum,kauzmann2013quantum} but are referred to as ``unicorns'' in chemistry because they cannot be physically observed~\cite{frenking2007unicorns}.

Aromaticity is one of the most ancient unicorns in chemistry~\cite{frenking2007unicorns}, tracing its origins back even further than the discovery of quantum superposition principle~\cite{kekule1865constitution,martin2015challenges,sola2016aromaticity}. However, it still lacks a comprehensive and universally accepted definition \cite{sola2017aromaticity,merino2023aromaticity}. Conventionally, both aromaticity and its counterpart, antiaromaticity, are believed to arise from electron delocalization within a group of atoms that form a closed loop, either in two or three dimensions~\cite{minkin1999glossary,chen2005nucleus, Sola2022}. In aromatic molecules, electron delocalization exhibits a homogeneous distribution, whereas in antiaromatic molecules, they display an inhomogeneous behavior due to localization of bonds~\cite {Feixas2010}. In more basic terms, a molecule is considered nonaromatic if its electronic structure cannot be represented by a quantum superposition that spreads over a closed region or volume in space.

This letter investigates the delocalization of electrons in the $\pi$-electronic structures of some archetypal monocyclic aromatic molecules as an application of RTS. To this aim, we consider the ground state of molecules in the basis of localized nonorthogonal atomic orbitals at the post-Hartree-Fock level of theory. By using the genuine superposition measure that we proposed in~\cite{pusuluk2022unified}, we show that the amount of superposition shared between biorthogonal atomic orbitals can effectively capture the aromaticity order of the molecules, comparable to the most successful measures of aromaticity used in the literature.

The structure of the paper is outlined as follows. We begin by providing an overview of quantifying superposition in nonorthogonal quantum states and its limitations, while in the following, we introduce the biorthogonal framework that addresses these limitations. Subsequently, we present results for applying these concepts to analyze electron delocalization in the $\pi$-electronic structures of ground states for some aromatic molecules.

%***==================================================***%
%  Quantifying superposition in nonorthogonal states  %
%***==================================================***%

\textit{Superposition in a nonorthogonal basis.}--- Consider a $d$-dimensional Hilbert space denoted by \(\mathcal{H}\), which possesses a normalized, linearly independent, and nonorthogonal basis consisting of states $|c_i\rangle$ such that \(\langle c_i|c_j\rangle = S_{ij} \in \mathbb{C}\). Any density operator $\hat{\rho}$ existing within \(\mathcal{H}\) can be represented using this basis as follows:
\begin{eqnarray} \label{eq1}
    \hat{\rho} = \sum_{i,j=1}^d \rho_{ij}|c_i\rangle\langle c_j|.
\end{eqnarray}
Here, the complex coefficients \(\rho_{ij}\) are equal to \(\langle c_i^\perp | \hat{\rho} | c_j^\perp \rangle\), where \(\{|c_i^\perp\rangle\}\) with \(\langle c_i^\perp |c_j \rangle = \delta_{i,j}\) is another nonorthogonal basis called the \textit{dual} of the basis \(\{|c_i\rangle\}\). 

When the density operator takes the form $\hat{\rho}_f=\sum_i p_i |c_i\rangle\langle c_i|$, where $p_i$ represents a probability distribution, the matrix $\rho$ formed by \(\rho_{ij}\) contains non-zero elements only in its diagonal. According to Refs.~\cite{theurer2017resource, torun2021resource, csenyacsa2022golden}, these states are referred to as superposition-free. Conversely, states that possess non-zero off-diagonal elements in $\rho$ are termed superposition states. By taking \(\hat{\rho}_f\) as a reference point or benchmark for quantifying and comparing the resource content of other states, the amount of superposition present in state~\eqref{eq1} can be quantified by
\begin{eqnarray} \label{eq2}
    l_1[ \rho ] = \sum_{i\neq j} |\rho_{ij}|.
\end{eqnarray}

However, as demonstrated in Ref.~\cite{pusuluk2022unified}, when we regard $\hat{\rho}_f$ as the state of a composite system, it becomes evident that it should encompass superposition. This superposition arises not from a mere linear combination of nonorthogonal states but rather from their mutual overlaps. Consequently, the off-diagonal elements of $\rho$ do not fully encode the complete information regarding the quantum superposition carried by the state $\hat{\rho}$.

%***==================================================***%
%  Quantifying superposition in biorthogonal states  %
%***==================================================***%

\textit{Superposition in a biorthogonal basis.}--- The limitations of the \emph{nonorthogonal} matrix representation $\rho$ become apparent when it fails to satisfy the unit-trace requirement. We should introduce the overlap or Gram matrix $S$, which contains the necessary overlap information to address this. Then, the unit-trace requirement can be fulfilled by considering the following expressions:
\begin{eqnarray} \label{eq3}
\text{tr}[ \hat{\rho} ]= \sum_i \langle c_i^{\perp}| \hat{\rho} |c_i\rangle = 1 = \text{tr}[ \rho \, S ].
\end{eqnarray}

On the other hand, an extension of the nonorthogonal basis $\{|c_i\rangle\}$ to $\{|c_i\rangle, |c_i^{\perp}\rangle \}$ naturally incorporates the overlap information through the structure of the dual basis~\cite{brody2013biorthogonal}, defined by
\begin{eqnarray} \label{eq4}
|c_i^{\perp}\rangle = \sum_j S_{ji}^{-1}|c_j\rangle.
\end{eqnarray}

By employing the biorthogonal basis $\{|c_i\rangle, |c_i^{\perp}\rangle \}$, we can represent the state described in Eq.~\eqref{eq1} as follows:
\begin{eqnarray} \label{eq5}
\hat{\rho} = \sum_{i,j=1}^d \bar{\rho}_{ij} |c_i^{\perp}\rangle \langle c_j|,
\end{eqnarray}
which allows us to construct an alternative matrix form for $\hat{\rho}$, known as the \emph{biorthogonal} matrix representation. It is composed of the elements $\bar{\rho}_{ij}=\langle c_i^{\perp}|\hat{\rho}|c_j\rangle$ and is denoted as $\rho{\textit{\tiny{BO}}}$. Although this matrix is non-Hermitian, it satisfies the unit-trace property. In other words, since $\rho{\textit{\tiny{BO}}} = \rho \, S$, the trace of $\rho{\textit{\tiny{BO}}}$ should be equal to one, as shown in~\eqref{eq3}. Therefore, $\rho{\textit{\tiny{BO}}}$ encapsulates all the (non-)probabilistic information in its (off-)diagonal elements. Consequently, we can quantify the total superposition in the state $\hat{\rho}$ using the following expression:
\begin{eqnarray} \label{eq6}
l_1[ \rho_{\textit{\tiny{BO}}} ] = \sum_{i\neq j} |\bar{\rho}_{ij}|.
\end{eqnarray}

Hence Eq. (\ref{eq6}) serves as a measure for genuine quantum superposition since it quantifies all types of superposition. On the other hand, Eq. (\ref{eq2}) only quantifies inter-basis superposition. It is worth noting that the biorthogonal framework presented here also facilitates consistent generalizations of conventional quantum mechanics for finite-dimensional systems~\cite{mostafazadeh2010pseudo,brody2013biorthogonal,ju2019non}. These generalizations incorporate non-Hermitian observables with complete, yet non-orthogonal, eigenstates.

\begin{table} [b]
\caption{Quantifying the quantum delocalization in the ground state of nonaromatic cyclopropene and its ions with the same triangular structure. According to both measures $l_1[\rho]$ and $l_1[\rho_{\textit{\tiny{BO}}}]$, the nonaromatic molecule exhibits the minimum superposition. The antiaromatic cyclopropenyl anion demonstrates the maximum inter-base superposition, while the aromatic cyclopropenyl cation displays the maximum genuine superposition. \label{tab:3MR} }
\begin{ruledtabular}
\begin{tabular}[t]{c | ccc}
 \\[-0.75em]
&  \includegraphics[width=0.055\textwidth]{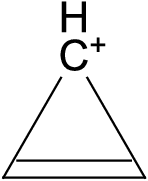} &   \includegraphics[width=0.055\textwidth]{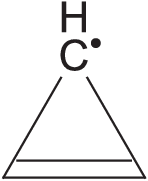} & \includegraphics[width=0.055\textwidth]{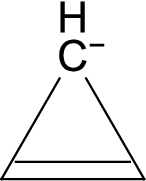}     \\[0.25em]
&  \makecell{Cyclopropenyl \\ Cation} & \makecell{Cyclopropenyl \\ Radical} & \makecell{Cyclopropenyl \\ Anion} \\ [0.05em]
\hline \\ [-0.35em]
$l_1[ \rho ]$ & 3.60 & 3.66 & 4.43 \\ [0.5em]
\hline \\ [-0.35em]
 $l_1[ \rho_{\textit{\tiny{BO}}} ]$ & 7.74 & 6.34 & 6.71 \\ [0.25em]
\end{tabular}
\end{ruledtabular}
\end{table}

\begin{table*} 
\caption{ \label{tab:5MR} The ground state delocalization of electrons within the \(\pi\) structure of five-membered heterocyclic molecules, represented as C\(_4\)H\(_4\)X, is quantified by superposition measures. Both measures $l_1[\rho]$ and $l_1[\rho_{\textit{\tiny{BO}}}]$ reliably provide the expected distinction between aromatic (X = \{CH\(^-\), NH, O\}) or antiaromatic (X = \{BH, CH\(^+\)\}) molecules.}
\begin{tabular}[t]{c | ccccc}
\hline \hline
 \\[-0.5em]
&   \includegraphics[width=0.085\textwidth]{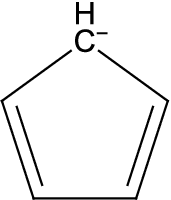} \hspace{2mm} &  \includegraphics[width=0.085\textwidth]{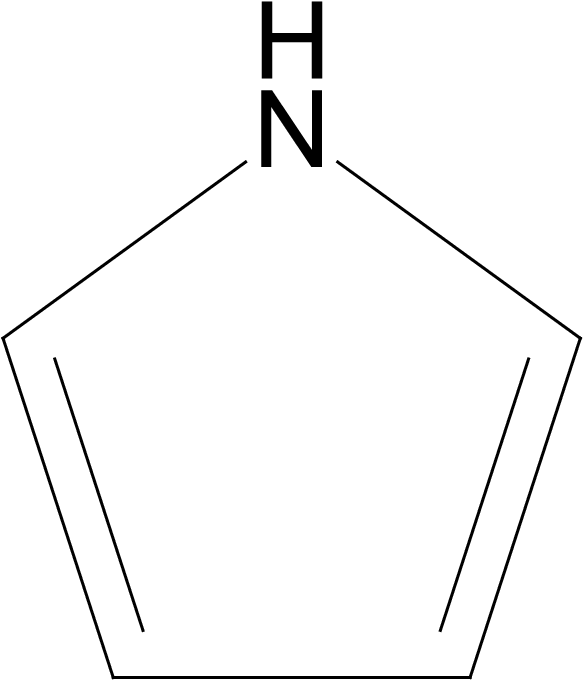} \hspace{2mm} &  \includegraphics[width=0.085\textwidth]{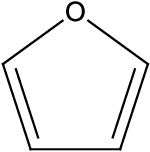} \hspace{2mm} &  \includegraphics[width=0.085\textwidth]{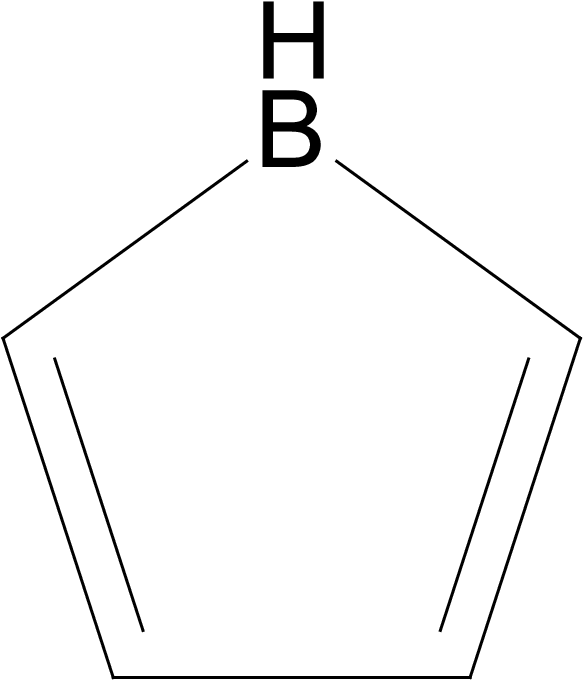} &  \includegraphics[width=0.085\textwidth]{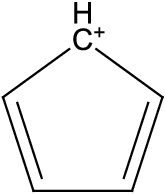}    \\[0.25em]
& \makecell{Cyclopentadienyl \\ Anion} \hspace{2mm} & \makecell{Pyrrole} \hspace{2mm} & \makecell{Furan} \hspace{2mm} & \makecell{Borole} & \makecell{Cyclopentadienyl \\ Cation} \\ [0.05em]
\hline \\ [-0.35em]
$l_1[ \rho ]$  &  26.48 \hspace{2mm} &  23.70 \hspace{2mm} &  22.93 \hspace{2mm} & 13.30 & 17.79  \\[0.5em]
\hline \\ [-0.35em]
$l_1[ \rho_{\textit{\tiny{BO}}} ]$ &  78.02 \hspace{2mm} & 68.93 \hspace{2mm} & 64.24 \hspace{2mm} & 39.53 & 49.14 \\  [0.5em] 
\hline \hline
\end{tabular}
\end{table*}

%***===========***%
%  Results  %
%***===========***%

\textit{Superposition in aromaticity}--- Let us explore the application of the superposition measures discussed thus far to analyze the quantum nature of chemical bonding phenomena. Specifically, our focus will be on examining some representative aromatic and antiaromatic molecules presented in Tables~\ref{tab:3MR},~\ref{tab:5MR},~and~\ref{tab:6MR}. These molecules exhibit planar and cyclic structures, wherein the electrons involved in the formation of $\pi$-bonds become delocalized across the atoms within the ring. This electron delocalization greatly affects the stability of the entire molecule and gives rise to various aromatic or antiaromatic properties. We can conceptualize it as a superposition that takes place within the cyclic $\pi$-subspace of the molecular electronic system. At this point, the question arises: what is the nature of this superposition? Does it correspond to inter-basis superposition, or is it a genuine superposition?

The electronic ground state of the molecules under investigation can be described as a linear combination of nonorthogonal states known as Slater determinants. These determinants are constructed using the occupation of \(p_z\) atomic orbitals (AOs) that are arranged circularly and form the $\pi$-subspace of the molecular electronic system (for additional information, please refer to Appendix \ref{app MO-CI}). The overlaps between the AOs solely determine the overlaps between the Slater determinants. Similarly, the duals of the Slater determinants are formed by the duals of the AOs. In the field of quantum chemistry, these AOs and their duals are commonly referred to as biorthogonal orbitals~\cite{1971_AOs, 1973_AOs, 1975_AOs, 1986_AOs}. In this context, the genuine quantum superposition present in the molecular electronic state can be understood as the quantum superposition shared between the biorthogonal AOs.

Today, we recognize the existence of various types of aromaticity. Among the oldest and simplest is H\"{u}ckel-type aromaticity. According to H\"{u}ckel rule, monocyclic conjugated hydrocarbons, also known as annulenes (C\(_n\)H\(_n\)), with D\(_{nh}\) symmetry are deemed aromatic if they contain \(4 n + 2\) \(\pi\)-electrons. Conversely, those with \(4 n\) \(\pi\)-electrons are classified as antiaromatic. The triangular structures in Table~\ref{tab:3MR} are the smallest molecular systems that follow this rule. According to our results, when cyclopropenyl radical (C\(_3\)H\(_3\)) is converted to either the aromatic or antiaromatic ionic forms, the amount of superposition possessed by the molecule increases. While the increase in the inter-basis superposition shared between nonorthogonal Slater determinants is more pronounced in the antiaromatic case (C\(_3\)H\(_3^-\)), the enhancement in aromaticity (C\(_3\)H\(_3^+\)) is more evident in the genuine superposition shared between biorthogonal Slater determinants.

The following molecules under consideration are pentagonal structures represented by C\(_4\)H\(_4\)X (see Table~\ref{tab:5MR}). The $\pi$-electronic structures of the cyclopentadienyl anion (C\(_5\)H\(_5^-\)), pyrrole (C\(_4\)H\(_5\)N), and furan (C\(_4\)H\(_4\)O) consist of 5 orbitals and 6 electrons. Based on H\"{u}ckel's rule, the cyclopentadienyl anion is considered aromatic. Also,  pyrrole and furan are known as hetero-aromatic molecules. The expected order of aromaticity among these three molecules is cyclopentadienyl anion $>$ pyrrole $>$ furan~\cite{sola2022aromaticity}.  Likewise, the cyclopentadienyl cation (C\(_5\)H\(_5^+\)) is considered antiaromatic according to H\"{u}ckel's rule. Borole (C\(_4\)H\(_5\)B), which also has 5 orbitals and 4 electrons in its $\pi$-electronic structure, exhibits less antiaromaticity compared to the cyclopentadienyl cation~\cite{sola2022aromaticity}. Our findings indicate that both $l_1[\rho]$ and $l_1[\rho_{\textit{\tiny{BO}}}]$ effectively quantify the extent of electron delocalization in the $\pi$-electronic structures of these five-membered molecules, consistent with the expected order of aromaticity and antiaromaticity among them. 

\begin{table*} [t]
 \caption{ \label{tab:6MR} Quantifying ground state quantum delocalization in six-membered ring molecules: Comparative analysis of quantum superposition measures $l_1[\rho]$ and $l_1[\rho_{\textit{\tiny{BO}}}]$ for assessing aromaticity order. Our results highlight the superior efficacy of $l_1[\rho_{\textit{\tiny{BO}}}]$ in evaluating electron delocalization within the $\pi$-structure of the visualized molecular systems compared to $l_1[\rho]$. }
\begin{ruledtabular}
\begin{tabular}[t]{c | cccccccc}
 \\[-0.75em]
 &  \includegraphics[width=0.085\textwidth]{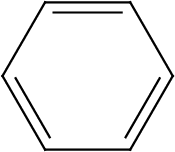} & \includegraphics[width=0.085\textwidth]{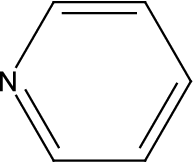} & \includegraphics[width=0.085\textwidth]{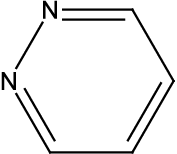} & \includegraphics[width=0.085\textwidth]{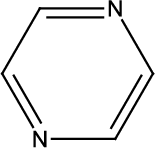} & \includegraphics[width=0.085\textwidth]{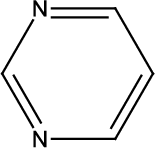} & \includegraphics[width=0.085\textwidth]{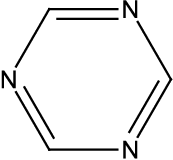} & \includegraphics[width=0.085\textwidth]{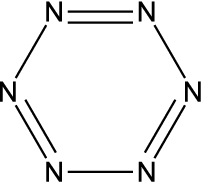} & \includegraphics[width=0.085\textwidth]{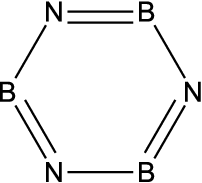} \\[0.45em]
 & Benzene & Pyridine & Pyridazine & Pyrazine & Pyrimidine & Triazine & Hexazine & Borazine \\ [0.35em]
 \hline \\ [-0.35em]
 $l_1[ \rho ]$ & 65.42 & 66.51 & 68.45 & 68.20 & 67.59 & 82.72 & 78.48 & 32.19  \\ [0.5em]
 \hline \\ [-0.35em]
 $l_1[ \rho_{\textit{\tiny{BO}}} ]$ & 267.04 & 265.96 & 263.67 & 265.11 & 264.32 & 250.94 & 255.47 & 133.32 \\  
\end{tabular}
\end{ruledtabular}
\end{table*}

Now, let us move on to a collection of molecules presented in Table~\ref{tab:6MR} and characterized by a hexagonal $\pi$-electronic structure comprising 6 orbitals and 6 electrons. These molecules are commonly classified as aromatic in the literature, with their expected order of aromaticity as follows: benzene (C\(_6\)H\(_6\)) $>$ pyridine (C\(_5\)H\(_5\)N)  $>$ pyridazine (C\(_4\)H\(_4\)N\(_2\)) $\approx$ pyrazine (C\(_4\)H\(_4\)N\(_2\)) $\approx$ pyrimidine (C\(_4\)H\(_4\)N\(_2\)) $>$ triazine (C\(_3\)H\(_3\)N\(_3\)) $>$ hexazine (N\(_6\)) $>$ borazine (B\(_3\)N\(_3\)H\(_6\)) \cite{cyranski2005energetic}. 

Evaluating the order above serves as one of the tests used to assess the effectiveness of electron delocalization-based aromaticity descriptors. However, even the most successful descriptors fail to accurately predict the complete expected trend. For instance, the \emph{para}-delocalization index (PDI) considers pyrazine, pyridazine, and hexazine more aromatic than benzene~\cite{sola2022aromaticity}, which contradicts the convention stating that as carbon atoms in benzene are replaced with nitrogen atoms, the aromaticity should decrease gradually. Similarly, the multicenter indices such as multicenter electron delocalization index (MCI) and I\(_\text{ring}\) infer that pyridazine is more aromatic than pyridine~\cite{feixas2008performance}.

Table~\ref{tab:6MR} demonstrates that the inter-basis superposition measured by $l_1[\rho]$ fails to replicate the expected aromaticity order. In contrast, the genuine quantum superposition quantified by $l_1[\rho_{\textit{\tiny{BO}}}]$ achieves a high degree of accuracy in reproducing this order, except for the ordering between triazine and hexazine. This finding suggests that the density matrix representation in the biorthogonal AO basis, $\rho_{\textit{\tiny{BO}}}$, encompasses comprehensive information about the extent of electron delocalization in the $\pi$-structure of an aromatic molecular system.

Besides, the deviation of our results from the conventional order, in which the triazine's $\pi$-electronic structure is expected to be more delocalized than those in hexazine, warrants further investigation. We previously mentioned that hexazine appears to be more aromatic than triazine, according to certain aromaticity measures like PDI. Our calculations may yield a similar finding due to the fact that, despite having more nitrogen atoms, hexazine possesses a more symmetric structure compared to triazine. In a sense, when we examine the overlaps between Slater determinants, we may measure not only the amount of electron delocalization but also the uniformity of its distribution. This observation is supported by the ordering of pyrazine \(>\) pyrimidine \(>\) pyridazine, as indicated in Table~\ref{tab:6MR}. Although these three molecules have an equal number of nitrogen atoms, the hydrocarbon pathways connecting the two nitrogen atoms exhibit symmetry in pyrazine, while they display complete antisymmetry in pyridazine.

Please note that the method we are proposing here is not restricted to planar molecules characterized by an electronic structure featuring a separation of \(\sigma\) and \(\pi\) orbitals. Our approach can be applied to any molecular system where the delocalization of electrons is known to occur within specific molecular orbitals. Thus, by utilizing the relevant orbitals, our method can effectively be employed.

%***===========***%
%   Conclusions   %
%***===========***%

\textit{Outlook.}--- Throughout this study, we have explored the concept of aromaticity as an application of the RTS. Our primary objective was to quantify the electron delocalization in the ground state $\pi$-electronic structures of representative monocyclic molecules regarding the quantum superposition present among localized AOs. We performed electronic state calculations using the CASSCF method and adopted the mode picture, which is particularly well-suited for characterizing quantum correlations in systems comprising indistinguishable fermionic particles~\cite{wiseman2003entanglement,benatti2020entanglement}. Furthermore, we utilized the \(l_1\) norm to measure the amount of quantum superposition in the electronic density matrices expressed in both nonorthogonal and biorthogonal AO bases. This comprehensive framework enabled us to capture not only the shared quantum superposition between nonorthogonal Slater determinants but also the local quantum superposition embedded within their overlaps.

Our study provides compelling evidence for accurately reproducing the expected order of aromaticity in most of the molecules we examined. This remarkable achievement is made possible through the genuine quantum superposition displayed by the biorthogonal density matrices. In contrast, the nonorthogonal density matrices, which carry inter-base quantum superposition, prove inadequate in accurately ordering the same molecules. These findings highlight the exceptional success of the biorthogonal framework in capturing all nonclassicality in nonorthogonal systems, as demonstrated by the electron delocalization in molecular systems. Moving forward, our objective is to replicate these calculations using valence bond structures as the nonorthogonal basis. In particular, the Kekul\'{e} basis~\cite{Douglas2003} holds great potential for investigating significantly larger molecules without compromising accuracy.

Additionally, our research uncovers the potential applications of chemical systems within the RTS, particularly in technological domains. Numerous experimental studies have already explored the phenomenon of delocalization in these systems~\cite{Hornberger2012, Brand2018, Brand2020}, providing a foundation for employing similar experimental setups to harness the power of superposition in various areas. This opens up exciting possibilities for leveraging molecular systems in cutting-edge fields such as quantum information, quantum metrology, and quantum cryptology. 

The biorthogonal extension of the RTS offers a comprehensive framework for understanding electron delocalization in the \(\pi\)-structure of molecular systems. However, the current superposition measures used in this study fail to distinguish between electron delocalization in aromatic and antiaromatic molecules. Previous research suggests that the distinction between aromatic and antiaromatic molecules lies not in the overall extent of electron delocalization but rather in the distribution of contributions from different bipartite delocalization terms~\cite{Feixas2010}. Although we have not explicitly addressed bipartite delocalization results here, our findings align with this perspective.

Additionally, different types of superpositions can be transformed into different forms of energy, as observed in the field of quantum thermodynamics~\cite{2016_Entropy_Ozgur, Pusuluk_2021_PRR}. We conjecture that the relationship between energy and superpositions in aromatic, nonaromatic, and antiaromatic molecules differs in this context, resulting in distinct kinetic and thermodynamic consequences. Discovering a method to uncover this connection could potentially lead us to a superposition-based measure of aromaticity and facilitate the development of a quantum resource theory for this phenomenon.

%In this study, we present a novel method that offers significant advancements in anatomically investigating electron delocalization within aromatic and antiaromatic systems. In addition to examining the bipartite terms analyzed in a previous study~\cite{Feixas2010}, our delocalization calculations encompass multipartite terms as well. For instance, we consider delocalization involving three atoms, which cannot be solely expressed in terms of delocalization between two atoms. Furthermore, our method allows us to decompose the overall electron delocalization into contributions based on electron number. This enables us to compare different types of contributions, such as one-electron delocalization, two-electron delocalization, three-electron delocalization, and so on. By incorporating these additional aspects, our method may pave the way for a more comprehensive analysis of electron delocalization, shedding light on the intricate details of aromatic and antiaromatic molecules.

\textit{Acknowledgments}---We are grateful to Ersin Yurtsever, Douglas J. Klein, Markus Reiher, Miquel Sol\`{a}, and \"{O}zg\"{u}r E. M\"{u}stecapl{\i}o\u{g}lu for useful suggestions and extensive discussions. O.P. would like to thank Ali Y{\i}ld{\i}z, Vlatko Vedral, and Martin B. Plenio for encouraging the utilization of the biorthogonal framework in the resource theory of superposition.

\textit{Competing Interests}---The authors declare no conflict of interest.

\textit{Contributions}---All the authors conceived the idea, derived the technical results, discussed all stages of the project, and prepared the manuscript and figures.

\textit{Materials \& Correspondence}---Correspondence and requests for materials should be addressed to O.P. (email: onur.pusuluk@gmail.com) or M.H.Y (email: mahirhyesiller@gmail.com)
 \vspace{10pt}

%***==========***%
%  ACKNOWLEDGMENTS %
%***==========***%

%\textbf{Author contributions}

%\lipsum[1]

%\textbf{Additional information}

%\lipsum[1]

%\textbf{Competing financial interests}

%The authors declare no competing interests.

%***==========***%
%  References    %
%***==========***%

%\bibliography{references} 
%

\clearpage
\appendix

%%%%%%%%%%%%%%%%%%%%%%%%%%%
\section{Construction of Electronic States} \label{app MO-CI}

Consider a molecular system composed of $N$ molecular orbitals (MOs) and $n_e$ electrons.  The dominant configuration of the system, known as the Hartree-Fock (HF) state $|\Psi_{HF}\rangle$, can be expressed as follows:
\begin{eqnarray} \label{eq7}
|\Psi_{HF}\rangle &=& |\psi_{n_e}\rangle\wedge\dots\wedge|\psi_{2}\rangle\wedge|\psi_{1}\rangle \nonumber \\
&\equiv& |11\dots 10\dots 0 \rangle_{\psi_1 \psi_2 \dots\psi_{n_e}\psi_{n_e+1}\dots\psi_{2N}}.
\end{eqnarray}
In the equation above, each MO consists of two spin modes with opposite values, referred to as molecular spin-orbitals. For instance, $|\psi_{2\mu-1}\rangle$ and $|\psi_{2\mu}\rangle$ represent the up and down spin-orbitals, respectively, of the $\mu$th MO. Additionally, to ensure the anti-symmetric algebra of the Fock space, the joint electronic system is constructed using the wedge product ($\wedge$) rather than the tensor product ($\otimes$).

Each system configuration corresponds to a Slater determinant, which is an antisymmetric product of occupied molecular spin-orbitals. For example, $|\Psi_{HF}\rangle \equiv |\psi_1 \psi_2 \dots \psi_{n_e}|$ represents the occupied MOs in the HF state. In this study, Our specific focus lies on the $\pi$-electronic sub-structure, which is characterized solely by the $\pi$-MOs and the $\pi$-electrons occupying them. Therefore, all the relevant configurations can be obtained by exciting the $\pi$-electrons from the reference state $|\Psi_{HF}\rangle$ within the $\pi$-MOs. Subsequently, the electronic ground state of the $\pi$-structure, denoted as $|\Psi_\pi\rangle$, can be expressed as a linear combination of these configurations:
 \begin{eqnarray} \label{eq8}
    |\Psi_\pi\rangle = \sum_{\Vec{s}}\lambda_{\Vec{s}} \hspace{1mm} | s_1 s_2 \dots s_{2N_{\pi}} \rangle_{\psi_1^{\pi} \psi_2^{\pi} \dots \psi_{2N_{\pi}}^{\pi}}.
\end{eqnarray}
Here, $N_{\pi}$ and $n_e^{\pi}$ represent the number of $\pi$-MOs and $\pi$-electrons, respectively. The set $\{\psi_\mu^{\pi}\}$ denotes the $\pi$-molecular spin-orbital basis. Each $s_\mu\in \{0,1\}$, and the sum of $s_\mu$ from $\mu=1$ to $2N_{\pi}$ equals $n_e^{\pi}$. A specific electronic configuration is represented by the vector $\vec{s}=\{s_1,s_2,...,s_{2N_{\pi}}\}$. The coefficients $\lambda_{\vec{s}}$ correspond to the different configurations and are determined by selecting the $\pi$-MOs and $\pi$-electrons as the active space and optimizing the $\pi$-MOs within this space using the CASSCF method.

The subsequent step involves rewriting $|\Psi_\pi\rangle$ in the nonorthogonal AO basis. To accomplish this, we leverage the fact that for the planar cyclic molecules we consider, each $\pi$-molecular spin-orbital ${\psi_{\mu}^{\pi}}$ can be expressed as a linear combination of the $p$-atomic orbitals $\{p_\nu^\pi\}$ involved in the $\pi$-bonds, as shown below:
\begin{equation} \label{eq9}
\psi_\mu\ =\ \sum_{\nu}{\ C_{\nu\mu}}p_\nu^\pi.
\end{equation}
Here, $C_{\nu\mu}$ represents the contribution of the $\nu$th atom's $p$-AO $p_\nu^\pi$ to the $\mu$th $\pi$-MO. The AO coefficients of MOs can be obtained from a CASSCF calculation.

Once we have obtained the coefficients for all $\pi$-MOs in Eq.~\eqref{eq9}, we can construct the coefficient matrix $C$. This matrix contains the coefficients of each $p$-AO in each $\pi$-MO. By utilizing this matrix, we can represent each configuration in Eq.~\eqref{eq8} as a linear combination of configurations in the $p$-AO basis~\cite{hiberty1978expansion,shaik2007chemist}. Consequently, we can expand the $\pi$-electronic structure state in \eqref{eq8} using the $\pi$-AO basis $\{p_\nu^\pi\}$ in the following manner:
\begin{equation} \label{eq10}
|\Psi_\pi^{AO}\rangle = \sum_{\vec{s}} \lambda_{\vec{s}}^{AO} |s_1 s_2 \dots s_{2N_{\pi}}\rangle_{p_1^\pi p_2^\pi \dots p_{2N_{\pi}}^\pi},
\end{equation}
where $\chi_{2\nu-1}$ and $\chi_{2\nu}$ denote the up and down atomic spin-orbitals of the $\nu$th atomic orbital $p_\nu^\pi$, respectively.  $\lambda_{\vec{s}}^{AO}$ represents the expansion coefficient of the electronic state in the $p$-AO basis.

To quantify the inter-basis superposition, we utilize Eq.~\eqref{eq2}, specifically $l_1[|\Psi_\pi^{AO}\rangle\langle\Psi_\pi^{AO}|]$. To represent the $\pi$-electronic structure state in the biorthogonal $p$-AO basis, we begin by obtaining the overlap matrix $S$. The elements of $S$ indicate the overlaps of electron configurations in the AO basis, determined through the inner product of configurations:
\begin{equation} \label{eq11}
S_{km}=\langle k_1k_2\dots k_{2N_{\pi}} | m_1m_2\dots m_{2N_{\pi}}\rangle_{p_1^\pi p_2^\pi \dots p_{2N_{\pi}}^\pi}.
\end{equation}

Once the overlap matrix is obtained, it becomes straightforward to derive the biorthogonal density matrix representation $\rho_{\textit{\tiny{BO}}}$ for $|\Psi_\pi^{AO}\rangle$, using the relationship $\rho_{\textit{\tiny{BO}}}=|\Psi_\pi^{AO}\rangle\langle\Psi_\pi^{AO}| \, S$.

%%%%%%%%%%%%%%%%%%%%%%%%%%%
%\section{HF results} \label{app HF}

%%%%%%%%%%%%%%%%%%%%%%%%%%%
\section{Computational Details} \label{compdetail}

Except for hexazine $(N_6)$, borole ($C_4H_5B$), and cyclopentadienyl cation ($C_5H_5^{+}$), all the molecules' optimized geometries were obtained at the B3LYP/aug-cc-pVTZ from the CCCBDB database \cite{johnson2013nist}. The other geometries were optimized in the Gaussian 09 \cite{g09} at the same level of theory. All HF and CASSCF calculations have been performed using the PySCF program package \cite{sun2018pyscf,sun2020recent} with STO-6G minimal basis set.

In the CASSCF calculation of 6-MR compounds, each molecule's $\pi$-electronic structure consists of 6 $\pi$-molecular orbitals (MOs) –-3 of which possess bonding and 3 with anti-bonding character–- and 6 electrons within these orbitals. Similarly,  the AO basis for each molecule is composed of 6 $p$-AOs, with each contributing one from a distinct atom in the ring to the $\pi$-electronic structure.

In MO (AO) basis, the $\pi$-electronic structures of the cyclopentadienyl anion $(C_5H_5^-)$, pyrrole $(C_4H_5N)$, and furan $(C_4H_4O)$ consist of 5 $\pi$-MOs ($p$-AOs) and 6 electrons, whereas those of borole $(C_4H_5B)$ and cyclopentadienyl cation $(C_5H_5^+)$ comprise 5 $\pi$-MOs ($p$-AOs) and 4 electrons. In pyrrole (furan), each carbon atom contributes to the $\pi$-system with one $p$-AO and an electron. In contrast, its nitrogen (oxygen) that undergoes ${sp}^2$ hybridization and places a lone electron pair in a $p$-AO contributes to $\pi$-system with a $p$-AO and its 2 electrons. However, in borole, the contributed $p$-AO of the boron atom to the $\pi$-electronic system is empty.

\newpage

\end{document}